\documentclass[conference]{IEEEtran}
\IEEEoverridecommandlockouts
\usepackage{cite}
\usepackage{amsmath,amssymb,amsfonts}
\usepackage{algorithmic}
\usepackage{graphicx}
\usepackage{textcomp}
\usepackage{todonotes}
\usepackage{xcolor}
\def\BibTeX{{\rm B\kern-.05em{\sc i\kern-.025em b}\kern-.08em
    T\kern-.1667em\lower.7ex\hbox{E}\kern-.125emX}}

\usepackage{pgfplots}
\pgfplotsset{compat=1.18}
\usepackage{scsnowman}

\usepackage{booktabs}

\usepackage[hidelinks]{hyperref}

\usepackage{listings}

\newcommand{\cpp}[1]{\lstinline[language=c++,keywordstyle ={\color{azure}},morekeywords={co_return,co_yield,co_await,ensure_started,sync_wait,start_detached,any_sender}]{#1}}

\usepackage{siunitx}
\usepackage[official]{eurosym}
\ifCLASSOPTIONcompsoc
\usepackage[caption=false,font=normalsize,labelfont=sf,textfont=sf]{subfig}
\else
\usepackage[caption=false,font=footnotesize]{subfig}
\fi

\usepackage{todonotes}

\newcommand{\orcid}[1]{\href{https://orcid.org/#1}{\includegraphics[height=10pt]{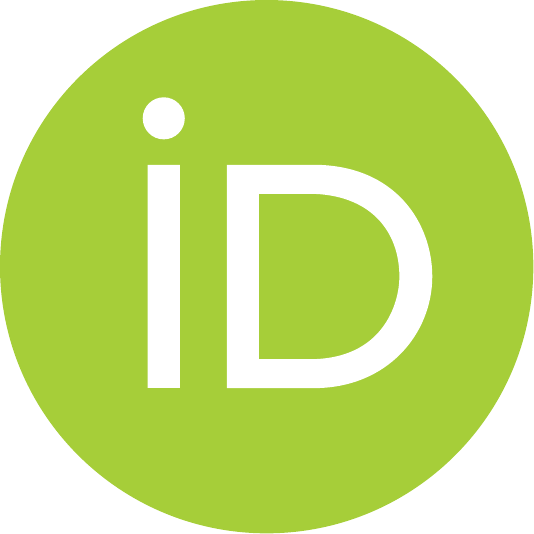}}}

\usepackage{xcolor} 
\definecolor{azure}{rgb}{0.0, 0.5, 1.0}
\definecolor{darkgreen}{rgb}{0.0, 0.5, 0.0}
\definecolor{amaranth}{rgb}{0.9, 0.17, 0.31}
\definecolor{cadetgrey}{rgb}{0.57, 0.64, 0.69}
\definecolor{aureolin}{rgb}{0.99, 0.93, 0.0}

\begin{document}

\title{Preparing for HPC on RISC-V: Examining Vectorization and Distributed Performance of an Astrophyiscs Application with HPX and Kokkos}


\author{
\IEEEauthorblockN{
Patrick Diehl\orcid{0000-0003-3922-8419}\IEEEauthorrefmark{7}\IEEEauthorrefmark{4}, 
Panagiotis Syskakis\orcid{0009-0005-0594-1445}\IEEEauthorrefmark{1}, 
Gregor Dai\ss\orcid{0000-0002-0989-5985}\IEEEauthorrefmark{2},
Steven R. Brandt\orcid{0000-0002-7979-2906}\IEEEauthorrefmark{1},
Alireza Kheirkhahan\orcid{0000-0002-4624-4647}\IEEEauthorrefmark{1}, \\ 
Srinivas Yadav Singanaboina\IEEEauthorrefmark{1}, 
Dominic Marcello\IEEEauthorrefmark{1},\\ 
Chris Taylor\orcid{0000-0001-7119-818X}\IEEEauthorrefmark{5}, 
John Leidel\IEEEauthorrefmark{5}, and Hartmut Kaiser\orcid{0000-0002-8712-2806}\IEEEauthorrefmark{1}\IEEEauthorrefmark{6}}
\IEEEauthorblockA{
\IEEEauthorrefmark{1}LSU Center for Computation \& Technology, Louisiana State University,
Baton Rouge, LA, 70803 U.S.A.}
\IEEEauthorblockA{\IEEEauthorrefmark{2} IPVS, University of Stuttgart,
Stuttgart, 70174 Stuttgart, Germany}
\IEEEauthorblockA{\IEEEauthorrefmark{4} Department of Physics and Astronomy, Louisiana State University,
Baton Rouge, LA, 70803 U.S.A.}
\IEEEauthorblockA{\IEEEauthorrefmark{5} Tactical Computing Labs, 1001 Pecan St.,
Lindsay, TX, 776250 U.S.A.}
\IEEEauthorblockA{\IEEEauthorrefmark{6} Department of Computer Science, Louisiana State University,
Baton Rouge, LA, 70803 U.S.A.}
\IEEEauthorblockA{\IEEEauthorrefmark{7} Applied Computer Science (CCS-7), Los Alamos National Laboratory, Los Alamos, NM 87545 U.S.A.\\
Email: diehlpk@lanl.gov}}

\maketitle

\begin{abstract}
In recent years, interest in RISC-V computing architectures has moved from academic to mainstream, especially in the field of High Performance Computing where energy limitations are increasingly a concern. As of this year, the first single board RISC-V CPUs implementing the finalized ratified vector specification are being released. The RISC-V vector specification follows in the tradition of vector processors found in the CDC STAR-100, the Cray-1, the Convex C-Series, and the NEC SX machines and accelerators. The family of vector processors offers support for variable-length array processing as opposed to the fixed-length processing functionality offered by SIMD. Vector processors offer opportunities to perform vector-chaining which allows temporary results to be used without the need to resolve memory references.

In this work, we use the Octo-Tiger multi-physics, multi-scale, 3D adaptive mesh refinement astrophysics application to study these early RISC-V chips with vector machine support. We report on our experience in porting this modern C\texttt{++} code (which is built upon several open-source libraries such as HPX and Kokkos) to RISC-V. In addition, we show the impact of the RISC-V Vector extension on a RISC-V single board computer by implementing the \cpp{std::experimental:simd} interface and integrating it with our code. We also compare the application's performance, scalability, and power consumption on desktop-grade RISC-V computer to an A64FX system.

The results presented in this paper are part of a longer-term evaluation of RISC-V's viability for HPC applications. 
\end{abstract}


\section{Introduction}
RISC-V is an instruction set architecture (ISA) for a general-purpose computer~\cite{waterman2014risc}. 
RISC-V is unique in that, besides its core instructions, the ISA offers a collection of extensions describing different features. One extension is the RISC-V ``V'' vector extension. The RISC-V extensions are composable, enabling hardware engineers to pick and choose the relevant parts for a desired machine design.

Unlike other ISA efforts, RISC-V's development is conducted in the open between hobbyists, academics, and commercial businesses that are working in a competitive market. Competition within the RISC-V community is left to the microarchitecture, hardware implementations of the ISA, possible custom ISA extensions, and system features (memory, storage, peripherals, etc) offered to consumers.

RISC-V's status as an open and free ISA has drawn international attention. The European Commission published a study on the impact of Open Source Software and hardware on the European Union's economy~\cite{blind2021impact}, and that included RISC-V. Partly because of this favorable review, the European Union (EU) announced the release of \euro{}270 million for building hardware and software based on the RISC-V instruction set. The European Processor Initiative (EPI)\footnote{\url{https://www.european-processor-initiative.eu/}}~\cite{kovavc2019european} is, perhaps, one of the more important projects for open source, RISC-V, and the development of low-power processors for extreme-scale computing. The project's ambitious aim is to have the first EU exascale system using RISC-V by 2026. As currently planned, the EU exascale computer will be a CPU only machine without acceleration cards. For these reasons, we decided to compare our RISC-V in-house cluster with supercomputer\ Fugaku which has Arm A64FX CPUs only.

In addition to the official recognition and promise of future support mentioned above, RISC-V is already making progress toward becoming an HPC technology. Several stable HPC software libraries have been successfully migrated, compiled, and tested using the development boards and single-board computers available so far. The following non-exhaustive list of HPC software has already been successfully migrated and evaluated on RISC-V: Fortran, clang, gcc, OpenMPI, OpenMP, OpenSHMEM, GASNet, UCX, and libfabric.

RISC-V machines up and until 2023 have consisted of IoT and single-board-computers. In December of 2023, a company called MILK-V released the first consumer desktop RISC-V computer, a machine they called ``Pioneer.'' In contrast to the single-board RISC-V computers, Pioneer features a SOPHON SG2042 RISC-V CPU. The SG2042 is an SoC with 64 cores, 64 MB of cache, and 128 GB of DDR4 on board memory. Previous consumer RISC-V processors have typically offered 4 cores along with 4 or 8 GB on board memory. The Pioneer's hardware configuration allows, for the first time, the possibility of making a real comparison of RISC-V with other HPC-grade hardware. In this particular case, the HPC-grade hardware is the A64FX nodes of the supercomputer\ Fugaku. Note that an A64FX node on Fugaku has 48 cores and 28 GB of HBM memory. Despite the desktop-grade CPU, the Milk-V computer only supports an early draft version (v0.7.1) of the RISC-V Vector (RVV) extension specification. For this reason, we also used the single board Banana Pi BPI-F3 for evaluating the vector extension. This board features a SpacemiT K1 RISC-V processor with a 256-bit vector unit which fully conforms to the RVV V1.0 specification. 
This paper is part of an ongoing evaluation of RISC-V's viability for HPC applications. Prior work~\cite{10.1145/3457388.3458657} demonstrated that a port of HPX to RISC-V had performance potential to be competitive with A64FX. In the previous paper, we used single-board computers with 4 cores and 8 GB memory. 
The more powerful CPU in this paper allows for larger problem size and a better comparison with the 48 cores of the A64FX CPU. Previously, we added the single-board computer's CPU as an architecture to Kokkos' CMake build system. For this paper, we had to add the SOPHON SG2042 RISC-V CPU. This paper continues the assessment of RISC-V by providing a focused comparison of a more sophisticated RISC-V machine with A64FX performance using a real-world astrophysics application: the Octo-Tiger HPX application. Octo-tiger is an astrophysics code capable of simulating collisions between stars. Octo-tiger uses a fast-multipole method over adaptive Octrees to evaluate the gravitational potential. Furthermore, we investigate the effect of vectorization on a single board RISC-V computer.

This study has made the following contributions to the HPC software stack:

\begin{itemize}
 \item A basic implementation of \lstinline{std::experimental::simd} for RISC-V, a library that implements SIMD types and operations in high-level C\texttt{++}.
  \item Kokkos support for MILK Pioneer's RISC-V CPU and SpacemiT K1 RISC-V CPU, see pull request \#6773\footnote{\url{https://github.com/kokkos/kokkos/pull/6773}} and \#7160\footnote{\url{https://github.com/kokkos/kokkos/pull/7160}}.
  \item Improved HPX support for locks on A64FX, see pull request \#6447\footnote{\label{note1}\url{https://github.com/STEllAR-GROUP/hpx/pull/6447}}. Initially, the target was A64FX, however, improvements on RISC-V were observed as well.
  \item The first performance study of desktop-grade RISC-V hardware supporting stellar merges using the astrophysics code Octo-Tiger.
  \item Providing an early performance study of the RISC-V Vector (RVV) extension on one of the first RISC-V CPUs with a fully RVV v1.0-compliant vectorization unit.
\end{itemize}

The paper is structured as follows: Section~\ref{sec:related:work} summarizes the related work. The software stack is introduced in Section~\ref{sec:software:stack}. The two scientific applications are briefly discussed in Section~\ref{sec::scientific:applications}. The in-house RISC-V cluster is presented in Section~\ref{sec:inhouse:cluster}. Node-level and distributed performance results are shown in Section~\ref{sec:performance}. The power consumption is analyzed in Section~\ref{sec:power}. Finally, the work is concluded in Section~\ref{sec:conclusion}.

%
%

\section{Related Work}
\label{sec:related:work}
Many vendors have provided RISC-V CPUs~\cite{10.1145/3457388.3458657}.
Most of the vendor provided CPUs are development boards, similar to the Raspberry Pi single-board computers for Arm CPUs. RISC-V single-board computers provide a foundation for the development and migration of software to RISC-V. The authors have performed a preliminary study~\cite{diehl2023evaluating} on these single-board computers using the Octo-Tiger software stack and have completed a scaling study on up to four cores. However, the Milk-V Pioneer's 64 core system allows us to extend that study.
Milk-V's Pioneer has been used for a comparison study with x86 CPUS and the \textit{RAJAPerf} benchmarking suite~\cite{hornung2017raja} in~\cite{10.1145/3624062.3624234}. A study on software support for academic and industrial applications is available here~\cite{9771410}. The following applications: cryptography~\cite{stoffelen2019efficient}, deep learning~\cite{louis2019towards}, and internet of things~\cite{schiavone2017slow} have been explored on RISC-V. However, we are not aware of any study of scientific simulation applications using RISC-V vector specification as of the time of this writing.

\section{Software stack}
\label{sec:software:stack}
In this work, we focus on the performance of our astrophysics application, Octo-Tiger, on RISC-V hardware. In this section, we present its software stack, features, and design decisions to provide a better context for the following results.

Octo-Tiger contains multiple, interleaved solvers (gravity, hydrodynamics and an experimental radiation solver) and uses adaptive mesh refinement (AMR).
While the AMR helps with the overall computational load, it makes it difficult to efficiently parallelize and distribute the work onto multiple compute nodes. 

In the following, we thus introduce both Octo-Tiger itself and the most important software frameworks it employs to alleviate this challenge and provide an efficient distributed implementation with portable compute kernels.
\subsection{HPX}
HPX is a distributed, asynchronous many-task runtime system~\cite{Kaiser2020} that allows a programmer to express data and execution dependencies by creating a task-graph 
through the use of futures and continuations.
Using this methodology, we can create millions of tasks, all managed by a backend which provides a single worker thread per core.
In the context of Octo-Tiger, we use these tasks for asynchronous tree-traversals. One thread can traverse the tree quickly, spawning additional parallel work as it executes, thus avoiding resource starvation.

HPX also provides various features to enable distributed computation:
It supports unified syntax and semantics for local and remote operations, asynchronous channels to exchange data, and is implemented for various networking backends (TCP, MPI, LCI~\cite{10.1145/3624062.3624598}).
These distributed features work together with HPX futures, allowing us to integrate communication into the task-graph, which, in turn, allows us to finely overlap computation and communication.
Octo-Tiger organizes information using an adaptive octree. The application of HPX's capabilities to the adaptive octree means we do not need to worry about whether a child/parent tree-node is on the same compute node when calling its methods. HPX takes care of moving the function call to the correct node.
These distributed features have been used in the past to run Octo-Tiger on machines such as Piz Daint~\cite{10.1145/3295500.3356221}, Summit~\cite{9556040} and, more recently, on the Supercomputer Fugaku~\cite{diehl-fugaku2024}. For current runs, we also target Perlmutter.

\subsection{Kokkos and HPX-Kokkos}
While HPX can help with the parallelization of the adaptive octree, there is still the issue of computational efficiency within the actual compute kernels that needs to be addressed. Specifically, improving efficiency by consecutive memory utilization and simple parallel-for patterns.

In order to address potential issues with computational efficiency, Octo-Tiger uses an entire sub-grid within each tree-node for efficiency.
This approach is a trade-off as it compromises the tree's adaptivity. Therefore, we usually limit the sub-grid size to $8x8x8$ (though this is adjustable at compile time).
Together with a ghost-layer for each sub-grid, these $8x8x8$ cells are the input for the actual compute kernels in the gravity and hydro solver.

This already eases the kernel development as the compute kernels deal with a regular, albeit small, data-structure. However, we still need to target various hardware platforms. Octo-Tiger needs to target both CPU and GPU supercomputers.

In order to address portability concerns, we use Kokkos. Kokkos is a performance portability framework, originally developed at
Sandia National Laboratories~\cite{9485033}.
Kokkos offers multiple hardware abstractions in the form of memory and execution spaces.
Using Kokkos means we only have to develop each compute kernel once, and then we can run them on GPUs made by any vendor.


Kokkos integrates with HPX which provides an additional incentive to adopt its use in Octo-Tiger.

Kokkos provides an HPX execution space~\cite{9460406}: This space allows users to execute a Kokkos kernel with HPX worker threads (eliminating the need for any conflicting thread pools that would occur when using the Kokkos OpenMP execution space).
Additionally, there is a HPX-Kokkos compatibility layer which allows programmers to make asynchronous Kokkos calls for most execution and memory spaces (works with the HPX, CUDA, HIP and SYCL spaces~\cite{10.1145/3585341.3585354}).
This compatibility layer allows direct integration of Kokkos kernels and data transfers into the HPX task graph.

Currently, all major compute kernels in Octo-Tiger's HPX hydro and gravity solver have been ported to Kokkos.
We use additional techniques to increase the efficiency given our specific use case within Octo-Tiger: For example, we use dynamic kernel fusion for the kernels on larger GPUs~\cite{daiss2022aggregation} (as the $8x8x8$ proved to be too much a bottleneck here otherwise, starving the GPU). 
To improve the efficiency on CPU execution spaces (including the HPX execution space), we furthermore use SIMD types within all major compute kernels. Kokkos already includes such SIMD types which we can use for this purpose. Using them allows us to use explicit vectorization (not relying on autovectorization at all) when executing a Kokkos kernel on the CPU. Yet, the same Kokkos kernel implementation stays compatible to the GPU as the SIMD data-types simply instantiate to scalar data-types on the GPU ensuring compatibility. On an x86 machine, the SIMD data-types instantiate to (for example) AVX512 types along with the appropriate AVX512 instructions.

Interestingly, with just minor changes one can also use SIMD types which implemenent the \lstinline{std::experimental::simd} interface, at least on the CPU execution spaces. In the past, we have used that to run our own (\lstinline{std::experimental::simd}-compatible) SVE types within the Octo-Tiger Kokkos kernels on A64Fx CPUs~\cite{daiss2022simd}. In that work we also showed Octo-Tiger's SIMD speedups on various other x86 CPU architecture, both when using the Kokkos SIMD types and when using the \lstinline{std::experimental::simd} types instead.

\subsection{RISC-V Vector (RVV) Library}
The compatibility to \lstinline{std::experimental::simd} types is also the key to make Octo-Tiger use the RISC-V vectorization. All we need are \lstinline{std::experimental::simd}-compatible types which use the RVV functionality underneath.



To assess RISC-V vectorization's impact on Octo-Tiger, we thus created a \lstinline{std::experimental::simd} backend for RISC-V, implementing all functionality required for Octo-Tiger.
The std::experimental::simd library is proposed for the C\texttt{++} Standard (SO/IEC TS 19570:2018) and provides a portable and platform-agnostic interface for SIMD types and operations. 

We implemented our backend\footnote{Available here: https://github.com/Pansysk75/cpp-simd-riscv} using the RISC-V Vector Intrinsics provided in GCC 14.
As Octo-Tiger can already utilize \lstinline{std::experimental::simd}~\cite{daiss2022simd}, the integration of this backend was a trivial process afterward.

This approach to SIMD may not always be as powerful as specialized kernels written in assembly code. However, there is significant speedup to be gained by utilizing a processor's SIMD capabilities. This library allows a programmer to explicitly take advantage of the SIMD unit without having to rely on the compiler's auto-vectorization feature. Furthermore, this approach allows us to have a single implementation for Kokkos kernels that works well on GPUs and still use explicit vectorization when being used on the CPU, simply by using the correct types for the given target device.

\subsection{Octo-Tiger}

Octo-Tiger is a specialized code designed to model self-gravitating astrophysical fluids, particularly focusing on interacting binary systems~\cite{10.1093/mnras/stab937}. Utilizing a combination of techniques, Octo-Tiger offers a comprehensive approach to understanding these complex astrophysical phenomena.

Octo-Tiger leverages several key features to accurately model interacting binary systems. Employing the finite volume method enables calculations of fluid dynamics. Octo-Tiger incorporates a rotating grid, aligning the frame of reference with the binary system's initial orbital period, reducing effects of numerical viscosity. Gravity modeling is facilitated through the Fast Multipole Method (FMM). Notably, the FMM conserves both linear and angular momentum, distinguishing it from conventional FMM approaches and enhancing simulation accuracy by enabling machine precision energy conservation within the rotating frame. Adaptive Mesh Refinement (AMR) techniques dynamically adjust computational grid resolution, allowing for efficient resource allocation by concentrating computational power where it is most necessary. Octo-Tiger evolves the system forward in time using the method of lines coupled with a third or Runge Kutta integrator. Additionally, Octo-Tiger's optimization for various architectures ensures that researchers can leverage its capabilities across diverse hardware platforms.

Due to its high optimization and the faster convergence of the finite volume method compared to smoothed particle hydrodynamics (SPH), Octo-Tiger has proven to be valuable for convergence studies (e.g. ~\cite{shiber2024hydrodynamic}, ~\cite{ diehletal2021}). These studies involve employing progressively finer resolutions on the same model to distinguish physical effects from numerical artifacts. 

\section{Scientific applications}
\label{sec::scientific:applications}
Having introduced Octo-Tiger, its software stack, and its features in the last section, we now take a look at the current research questions (beyond mere performance studies) for which Octo-Tiger is actively being used. Notably, we use some of these production simulations (for just a few time-steps at a low resolution) to generate the performance comparisons later on. Hence, we focus on the viability of the RISC-V hardware 
for a real-world production science code.

\subsection{Double White Dwarf Systems}

A double white dwarf (DWD) system consists of two compact stellar remnants, each having exhausted its nuclear fuel and collapsed to a dense, Earth-sized object. Over time, gravitational wave emission causes the orbital separation to decrease. If this becomes small enough, mass can be pulled from one white dwarf, the ``donor'', to the other, the ``accretor'', and in many cases this interaction is unstable, leading to runaway mass transfer and merger.

The potential for a Type Ia supernova arises if their combined mass exceeds 1.4 $M_\odot$. Conversely, if the interaction results in a merger with a combined mass below this threshold, it can lead to the formation of an intriguing celestial object known as R Coronae Borealis star (RCB). RCB stars are characterized by their peculiar elemental abundances, notably low to non-existent hydrogen content. Since white dwarfs are what is leftover after a star uses all of its hydrogen fuel, this suggests DWDs as a formation channel. While Octo-Tiger cannot model thermonuclear explosions like Type Ia supernova, it serves as a valuable tool for studying RCB formation. In particular, it can be used to study the dredge up of elements such as Oxygen-16 from the cores of the accretor to the surface of the RCB star during merger. We use restart files from production runs close to the merger~\cite{shiber2024hydrodynamic} where adaptive mesh refinement happened as one of our tests. In this case, the mesh is very unbalanced and shows characteristics of a production run.

We will run two types of tests using a double white dwarf: DWD Separated and DWD Merging. The former corresponds to the beginning of a typical scientific simulation. At this point, the grids are fairly regular and balanced and we run it for 10 time steps. The latter corresponds to a checkpoint restart from a time very near the merging of the two stars. At this point, the grids have had time to dynamically reshape themselves to the physics and are far less regular and balanced. We run this one for 20 time steps.

\subsection{v1309}
In contact binary systems, two stars orbit so closely that they share a common envelope of gas. As one or both stars evolve, they may reach a critical point where one star expands and engulfs its companion. This process can culminate in a luminous red nova (LRN), characterized by a sudden surge in brightness and the ejection of material. 

One notable example of such an event occurred in September 2008 when the contact binary star system V1309 Sco underwent a merger~\cite{Tylenda_2011}. The system's brightness increased by a factor of approximately one hundred. What made the observation of V1309 Sco remarkable was the extensive pre-merger data collected by the Optical Gravitational Lensing Experiment (OGLE). For six years leading up to the merger, OGLE meticulously documented the system's behavior, providing a rare opportunity to observe a stellar merger both before and after the event.

One goal of Octo-Tiger is to be able to model the light curve generated by merger events like V1309 Sco. This requires careful treatment of the transition between the optically thick and optically thin regions in the stellar atmosphere. Octo-Tiger is able to add extra resolution to the stellar atmosphere, providing a more accurate representation of the transition region. Octo-Tiger also has a specialized implementation of the self consistent field (SCF) module used to generate initial conditions for binary stars~\cite{Even_2009}. This version is able to produce models with common atmospheres, like the progenitor of V1309 Sco. We have included a model of V1309 Sco in this paper. This model does not include the effects of radiation transport, since this module of Octo-Tiger had not been developed at the time it was produced.

This test tends to be quite memory intensive and we run it for 10 time steps.

\section{In-House RISC-V Cluster}
\label{sec:inhouse:cluster}
\subsection{MILK-V Pioneer}
\begin{figure}[tb]
    \centering
    \includegraphics[width=0.5\linewidth]{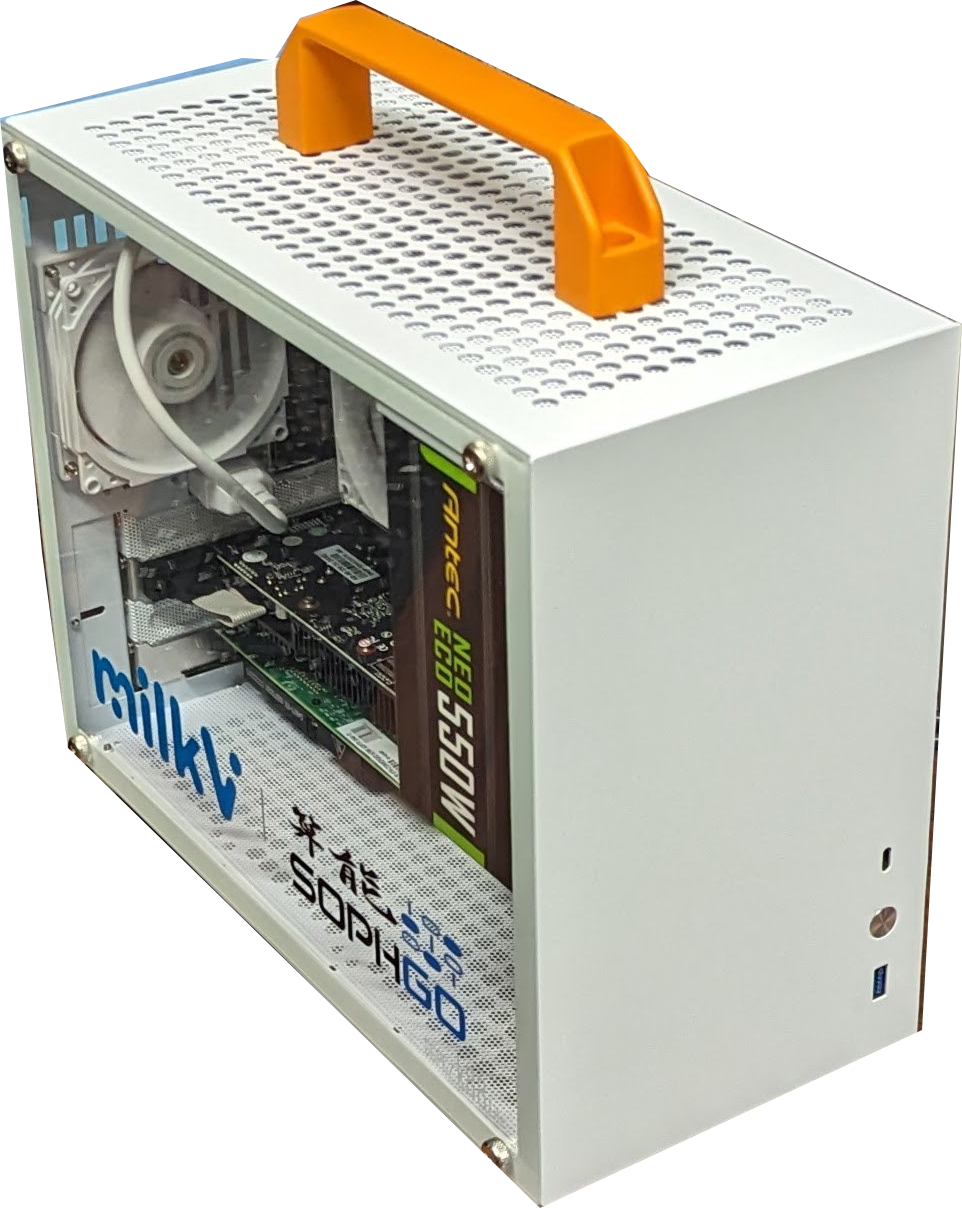}
    \caption{Image of one of the MILK-V Pioneer nodes of the in-house cluster. Each node has a 64-core SOPHON SG2042 RISC-V CPU and 128 GB DDR4 System Memory.}
    \label{fig:milk:v}
\end{figure}

For this paper, we built an in-house cluster with two nodes. Each node is a Milk-V Pioneer desktop machine with a 64-core SOPHON SG2042 RISC-V CPU, 128 GB DDR4 RAM, and a 1TB PCI 3.0 SSD card. The nodes are connected via Ethernet using Intel X540-T2 network cards. Figure~\ref{fig:milk:v} shows an image of one of the nodes. We named the cluster Olaf \scsnowman[arms, eyes, mouth, nose, buttons] after the snowman from the movie Frozen, since the color scheme reminded us of that character.

The machine comes with \emph{Fedora 38} preinstalled. The original disk partition occupied only a fraction of the NVMe disk, so we expanded it. After that, it was easy to install additional packages with \emph{dnf}. The default \emph{slurm} provided by the repository was sufficient to create a cluster from the two nodes. The system applications were installed on the local disk, the libraries and the test application as well as users' home directories were hosted by an NFS shared file system. We were not able to upgrade the Linux kernel provided by the repository, all attempts resulted in error messages that the new kernel is in conflict with installed kernel. Setting aside the kernel upgrade issue, the current kernel worked perfectly for our test environment. It should be noted that the need for security updates will require administrators to find a way to upgrade the Linux kernel. 

\subsection{Banana Pi BPI-F3 }
For experimenting with RISC-V Vectorization, we utilized the Banana Pi BPI-F3 development board, equipped with a SpacemiT K1 8-core RISC-V chip. This is a low-power processor on a board with just 4GB of memory. It runs Bianbu Linux, a Board Support Package provided by SpacemiT which is based on Debian Linux. The processor features 256-bit vectorization and is fully compliant with the V1.0 specification of the RISC-V Vector extension, which is starting to be supported by major compilers such as Clang and GCC.

\section{Performance results}
\label{sec:performance}
First, we show the effect of the scalable vector extensions on the Banana Pi BPI-F3 board, see Section~\ref{sec:banana:results}. We show node-level scaling for the astrophysics application Octo-Tiger using the rotating star test problem and the double white dwarf (DWD) real-world scenarios described in Section~\ref{sec:performance:node:level} on MILK-V. In Section~\ref{sec:dsitrbuted:runs}, we show distributed results using two real-world scenarios, namely DWD and v1309. Table~\ref{tab:software} shows the compiler and software versions used in this paper on the RISC-V cluster.
\begin{table}[tb]
\caption{Compiler and software versions used to build Octo-Tiger with CMake 3.19.5. We optimized HPX for RISC-V and A64FX and used a specific commit of HPX's development branch for some of the runs.}
    \label{tab:software}
    \centering
    \begin{tabular}{ccccc}\toprule
     gcc    & HPX & Boost & openmpi & hwloc  \\\midrule
    13.2.1   & \textit{72ca840}/1.9.1  & 1.84  & 4.1.2  & 1.11.12    \\\midrule
      Kokkos & HPX-Kokkos  & cppuddle & jemalloc & Octo-Tiger  \\
        4.0.01  & 0.4.0 &
\textit{c084385} & 5.2.1 & \textit{dd5cb880}
      \\\bottomrule
    \end{tabular}
\end{table}

\subsection{Effect of scalable vector extensions (Banana Pi BPI-F3) }
\label{sec:banana:results}

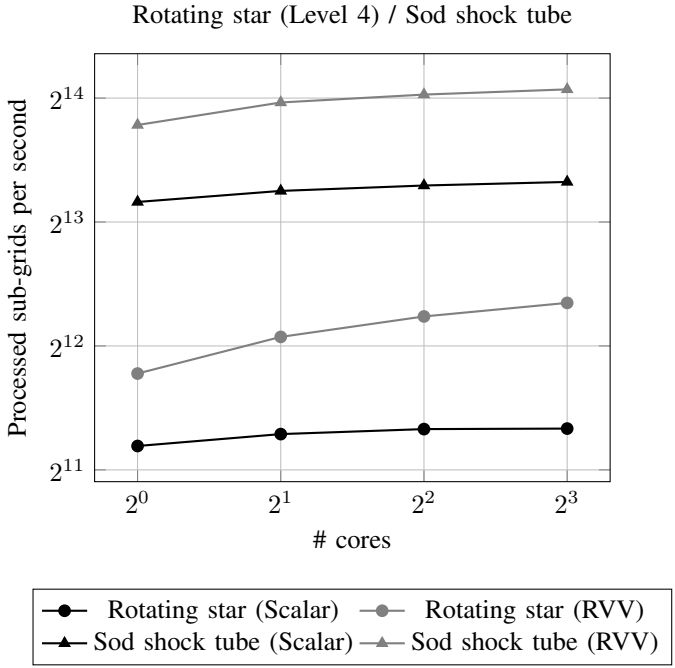
\begin{figure}[tb]
\centering
\begin{tikzpicture}
   \begin{axis}[xlabel={\# cores},ylabel={Processed sub-grids per second},title={Rotating star (Level 4) / Sod shock tube},grid,xmode=log,log basis x={2},xtick={1,2,4,8},ymode=log,log basis y={2},legend columns=2,legend style={at={(0.5,-.25)},anchor=north}]
    \addplot[thick,mark=*,black] table [x expr=\thisrowno{0},y expr={1184*512*25/\thisrowno{1}}, col sep=comma] {rotating_star_scalar.csv};
    \addplot[thick,mark=*,gray] table [x expr=\thisrowno{0},y expr={1184*512*25/\thisrowno{1}}, col sep=comma] {rotating_star_simd.csv};
    \addplot[thick,mark=triangle*,black] table [x expr=\thisrowno{0},y expr={8*512/\thisrowno{1}*25}, col sep=comma] {sod_scalar.csv};
    \addplot[thick,mark=triangle*,gray] table [x expr=\thisrowno{0},y expr={8*512/\thisrowno{1}*25}, col sep=comma] {sod_simd.csv};
    \legend{Rotating star (Scalar), Rotating star (RVV),
    Sod shock tube (Scalar),Sod shock tube (RVV)};
    \end{axis}
    \end{tikzpicture}
    \caption{Single node scaling for a rotating star on Banana Pi BPI-F3 using scalar values and RISC-V vector extensions. }
    \label{fig:single-node-rotating-star-sve}
\end{figure}

Due to the limited memory of 4 GB, we could only run some test problems and not the DWD scenario.
Figure~\ref{fig:single-node-rotating-star-sve} shows the scaling from a single core up to eight cores (lines with the star marker) for the sod shock tube (hydro only) example.
Additionally, the figure shows the rotating star (hydro and gravity solver) example with four mesh refinements (level 4) are the lines with a square marker. 
This was the largest adaptively refined mesh fitting within the memory of the single board computer.
For both scenarios, we observe scaling from one core to two cores but for more cores the scaling is marginal.
We have seen similar behavior on Rasberry Pis with ARM CPUs~\cite{gupta2020deploying}.

More interesting and noticeable is the effect of vectorization.
We see a factor of around $1.7$ in speedup for using vectorization for the sod shock tube example, and about $2$ in speedup for the rotating star example.
Considering the $256$ bit vector width and our usage of double precision, the optimal speedup would have been $4$, however, we only use the explicit vectorization with types within Octo-Tiger's compute kernels, not within the rest of the code (such as the tree management). Hence, the lower speedup.
Overall, we consider this a promising first result for our experiments with the RISC-V vectorization! However, it still falls a bit short of the speedup we experience when using the \lstinline{std::experimental::simd} types on AVX2 platforms, where we achieved a speedup of around $2.6$ in \cite{daiss2022simd} using similar Octo-Tiger scenarios.

\subsection{Nodel-level scaling (MILK-V)}
\label{sec:performance:node:level}
\subsubsection{Rotating star}
\begin{figure}[tb]
\centering
\begin{tikzpicture}
   \begin{axis}[xlabel={\# cores},ylabel={Processed sub-grids per second},title={Rotating star (Level 5)},grid,xmode=log,log basis x={2},xtick={1,2,4,8,16,32,64},ymode=log,log basis y={2},legend columns=2,legend style={at={(0.5,-.25)},anchor=north}]
    \addplot[thick,mark=*,black] table [x expr=\thisrowno{0},y expr={5048*512/\thisrowno{1}*10}, col sep=comma] {rotating-level-5-cpu.csv};
    \addplot[thick,mark=*,gray] table [x expr=\thisrowno{0},y expr={5048*512/\thisrowno{1}*10}, col sep=comma] {rotating-level-5-cpu-lock.csv};
    \addplot[thick,mark=square*,black] table [x expr=\thisrowno{0},y expr={5048*512/\thisrowno{1}*10}, col sep=comma] {rotating-level-5-fugaku.csv};
    \addplot[thick,mark=square*,gray] table [x expr=\thisrowno{0},y expr={5048*512/\thisrowno{1}*10}, col sep=comma] {rotating-level-5-fugaku-lock.csv};
    \legend{MILK-V, MILK-V (optimized),A64FX, A64FX (optimized)};
    \end{axis}
    \end{tikzpicture}
    \caption{Single node scaling for a rotating star on ARM A64FX and RISC-V.}
    \label{fig:single-node-rotating-star}
\end{figure}
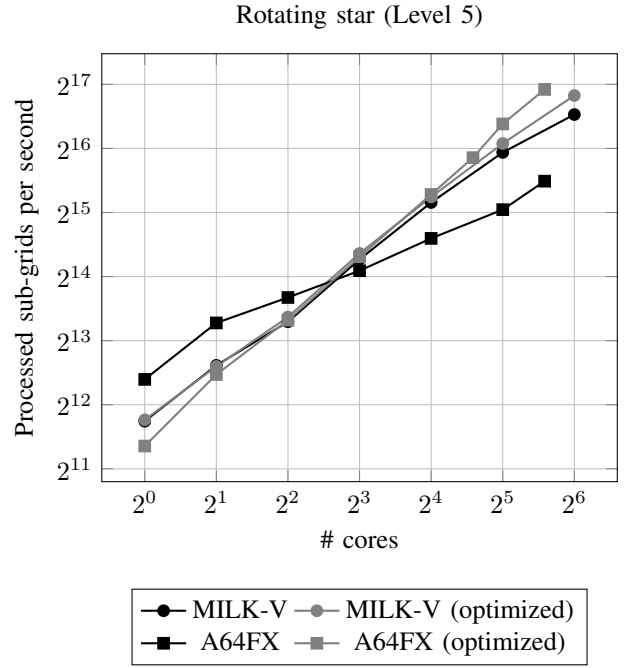

We use the single rotating star example from Octo-Tiger's test suite.
Figure~\ref{fig:single-node-rotating-star} shows the scaling for the level of refinement 5 with an octree containing 2,220 leaves and 2,584,576 cells.

We ran this scenario on a single MILK-V board with 64 cores in total and on a single Supercomputer\ Fugaku node with 48 cores. For simulations using 8 cores or less, A64FX was faster. At higher core counts, the RISC-V CPU (black lines) was faster. We investigated the poor performance on A64FX and discovered that 128-bit atomics were actually simulated using locks on both A64FX and RISC-V by the C\texttt{++} standard library. This had a significant impact on performance because HPX uses 128-bit atomics in its internal scheduling and synchronization.
An alternate solution for the schedulers is to use 64-bit atomics, which are lock-free on both A64FX and RISC-V. The 64-bit atomics are enabled for HPX schedulers by configuring \lstinline[language=c++]{HPX_LOCKFREE_PTR_COMPRESSION=ON} when invoking CMake, see pull request \# 6447\footref{note1}.

We executed the same test using the same dependencies, except for a newer HPX version with the optimization (gray lines). After optimizing, the performance of A64FX and RISC-V are almost identical (but now reversed, with A64FX being slightly faster above 8 cores).

\subsubsection{DWD}
This section uses the mesh from some recent production runs for a double white dwarf (DWD) merger. We use real-world production runs and initial data for these tests. In particular, we ran DWD Separated (where the stars are widely separated), and DWD Merging (which uses restart files to follow a nearly complete simulation) and ran with both 10 and 11 levels of adaptive mesh refinement~\cite{shiber2024hydrodynamic}. Table~\ref{tab:data:set} shows the average floating point operations measured using the \textit{perf} tool on an Intel Skylake CPU.

\begin{table}[tb]
    \centering
    \caption{Average floating point operations (FLOP) per timestep, Number of cells, the memory usage, and the file size of the input file for all three refinement levels.}
    \label{tab:data:set}
    \begin{tabular}{c|llll}\toprule
     Level & FLOP &  \# cells & Memory & File size   \\\midrule
     10 (initial) &  \num{5.28E+11}    &  1.6M      & 5 GB        &  72MB      \\
     10 (refined) & \num{1.68E+12}  &   3.8M &   11 GB &    548MB \\
     11 (initial)  &  \num{1.11E+12}                &  3.6M     &  10 GB       &  162MB       \\
     11 (refined) & \num{1.75E+13}  &  40.2M &  113 GB &  5.8 GB \\\bottomrule
    \end{tabular}
\end{table}

Figure~\ref{fig:dwd:single:beginning} shows the scaling for both runs. With 10 levels of refinement, the code has higher throughput because there are fewer cells. Both levels scale well on the single node. The gray line shows the runs that use atomics instead of locks. We see a significant improvement in this case for runs with 11 levels of refinement, but still see some improvement for 10 levels of refinement (but only for higher core counts). Figure~\ref{fig:dwd:single:refined} shows the DWD Merging experiment with 10 and 11 levels of refinement.

\begin{figure}[tb]
\centering
 \subfloat[\label{fig:dwd:single:beginning}]{
\resizebox {0.95\linewidth} {!} {
\begin{tikzpicture}
   \begin{axis}[xlabel={\# cores},ylabel={Processed sub-grids per second},title={DWD Separated},grid,xmode=log,log basis x={2},xtick={1,2,4,8,16,32,64},ymode=log,log basis y={2},legend columns=2,legend style={at={(0.5,-.25)},anchor=north},axis y line*=left]
   \addplot[thick,mark=*,black] table [x expr=\thisrowno{0},y expr={2220*512/\thisrowno{1}*20}, col sep=comma] {dwd-level-10-cpu.csv};
    \addplot[thick,mark=*,gray] table [x expr=\thisrowno{0},y expr={2220*512/\thisrowno{1}*20}, col sep=comma] {dwd-level-10-cpu-lock.csv};
    \addplot[thick,mark=square*,black] table [x expr=\thisrowno{0},y expr={4796*512/\thisrowno{1}*10}, col sep=comma] {dwd-level-11-cpu.csv};
    \addplot[thick,mark=square*,gray] table [x expr=\thisrowno{0},y expr={4796*512/\thisrowno{1}*10}, col sep=comma] {dwd-level-11-cpu-lock.csv};
    \legend{Level 10,Level 10 (optimized),Level 11, Level 11 (optimized)};
    \end{axis}
    \begin{axis}[hide x axis,axis y line*=right,xmode=log,log basis x={2},xtick={1,2,4,8,16,32,64},ymode=log,ylabel=GFLOP/s,ylabel near ticks,ytick={1.11e12/23724.2*10/1e9,5.28e11/5911.97*20/1e9,5.28e11/976.978*20/1e9,5.28e11/172.159*20/1e9},yticklabels={0.47,1.79,10.8,61.33}]
    \addplot[thick,mark=*,black,draw=none,forget plot] table [x expr=\thisrowno{0},y expr={5.28e11/\thisrowno{1}*20/1e9}, col sep=comma] {dwd-level-10-cpu.csv};
    \addplot[thick,mark=*,gray,draw=none,forget plot] table [x expr=\thisrowno{0},y expr={5.28e11/\thisrowno{1}*20/1e9}, col sep=comma] {dwd-level-10-cpu-lock.csv};
    \addplot[thick,mark=square*,black,draw=none,forget plot] table [x expr=\thisrowno{0},y expr={1.11e12/\thisrowno{1}*10/1e9}, col sep=comma] {dwd-level-11-cpu.csv};
    \addplot[thick,mark=square*,gray,draw=none,forget plot] table [x expr=\thisrowno{0},y expr={1.11e12/\thisrowno{1}*10/1e9}, col sep=comma] {dwd-level-11-cpu-lock.csv};
    \end{axis}   
    \end{tikzpicture}
    }}
    
   \subfloat[\label{fig:dwd:single:refined}]{
   \resizebox {0.95\linewidth} {!} {
   \begin{tikzpicture}
   \begin{axis}[xlabel={\# cores},ylabel={Processed sub-grids per second},title={DWD Merging},grid,xmode=log,log basis x={2},xtick={1,2,4,8,16,32,64},ymode=log,log basis y={2},legend columns=2,legend style={at={(0.5,-.25)},anchor=north}]
    \addplot[thick,mark=*,black] table [x expr=\thisrowno{0},y expr={7344*512/\thisrowno{1}*20}, col sep=comma] {dwd-level-10-cpu-close.csv};
    \addplot[thick,mark=*,gray] table [x expr=\thisrowno{0},y expr={7344*512/\thisrowno{1}*20}, col sep=comma] {dwd-level-10-cpu-close-lock.csv};
    \addplot[thick,mark=square*,gray] table [x expr=\thisrowno{0},y expr={78415*512/\thisrowno{1}*20}, col sep=comma] {dwd-level-11-cpu-close-lock.csv};
    \legend{Level 10,Level 10 (optimized), Level 11 (optimize)};
    \end{axis}
    \begin{axis}[hide x axis,axis y line*=right,xmode=log,log basis x={2},xtick={1,2,4,8,16,32,64},ymode=log,ylabel=GFLOP/s,ylabel near ticks,ytick={1.68e12/23724.2*20/1e9,1.68e12/4017.47*20/1e9,1.68e12/686.987*20/1e9},yticklabels={1.42,8.36,48.99}]
    \addplot[thick,mark=*,black] table [x expr=\thisrowno{0},y expr={1.68e12/\thisrowno{1}*20/1e9}, col sep=comma] {dwd-level-10-cpu-close.csv};
    \addplot[thick,mark=*,gray] table [x expr=\thisrowno{0},y expr={1.68e12/\thisrowno{1}*20/1e9}, col sep=comma] {dwd-level-10-cpu-close-lock.csv};
    \addplot[thick,mark=square*,gray] table [x expr=\thisrowno{0},y expr={1.75e13/\thisrowno{1}*20/1e9}, col sep=comma] {dwd-level-11-cpu-close-lock.csv};
    \end{axis}   
    \end{tikzpicture}
    }}
    \caption{Single node scaling for \protect\subref{fig:dwd:single:beginning} DWD Separated and \protect\subref{fig:dwd:single:refined} DWD Merging, respectively. For DWD Merging with 11 levels, we only used the optimized code. \textbf{The run on 8 cores took around 34 hours and we skipped the runs on 4 cores, 2 cores, and a single core since these runs were not feasible.} }
    \label{fig:dwd:single}
\end{figure}

\subsection{Distributed scaling (MILK-V)}
\label{sec:dsitrbuted:runs}

\subsubsection{DWD}
First, we executed the runs for the node-level scaling in the previous section using all 64 cores per node on two nodes. Figure~\ref{fig:performance:distributed} shows the results for level 10. For the DWD Separated in Figure~\ref{fig:performance:distributed:initial}, we observe some improvements, however, the workload was not large enough. For the DWD Merging case in Figure~\ref{fig:performance:distributed:refined}, the workload was sufficient and we observed a better improvement. For all node counts the A64FX runs were slower. However, we have to take into account that each A64FX node has 48 cores while each RISC-V node has 64 cores.

\begin{figure}[tb]
    \centering
\subfloat[\label{fig:performance:distributed:initial}]{
\begin{tikzpicture}[scale=0.95, transform shape]
 \begin{axis}[
    xbar=12pt,
    xmin=0,xmax=800,
    ytick=data,
    enlarge y limits={abs=1cm},
    yticklabels={1,1,2,2},
    ytick={1,2,3,4},
    bar width = 10pt,
    xlabel=Processed sub-grids per time step per second, 
    ytick align=outside, 
    ytick pos=left,
    ylabel= \# nodes,
    major x tick style ={transparent},
    legend style={at={(0.64,0.96)},anchor=north west, font=\footnotesize, legend cell align=left},
    xmajorgrids=true,
    nodes near coords,
    title=Level 10 (Separated)
    ]    
    \addplot[xbar,fill=cadetgrey!20, area legend] coordinates {
        (2220*512/172.159/20,1.25)
        (2220*512/134.226/20,3.25)
        };
        \addplot[xbar,fill=black!70, area legend] coordinates {
        (2220*512/282.168/25,1.75)
        (2220*512/249.388/25,3.75)
        };
        \legend {RISC-V, A64FX};
\end{axis}
\end{tikzpicture}
}

\subfloat[\label{fig:performance:distributed:refined}]{
\begin{tikzpicture}[scale=0.95, transform shape]
 \begin{axis}[
    xbar=12pt,
    xmin=0,xmax=800,
    ytick=data,
    enlarge y limits={abs=1cm},
    yticklabels={1,1,2,2},
    ytick={1,2,3,4},
    bar width = 10pt,
    xlabel=Processed sub-grids per time step per second, 
    ytick align=outside, 
    ytick pos=left,
    ylabel= \# nodes,
    major x tick style ={ transparent},
    legend style={at={(0.64,0.96)},anchor=north west, font=\footnotesize, legend cell align=left},
    xmajorgrids=true,
    nodes near coords,
    title=Level 10 (DWD Merging)
    ]    
    \addplot[xbar,fill=cadetgrey!20, area legend] coordinates {
        (7344*512/564.26/20,1.25)
        (7344*512/337.911/20,3.25)
        };
        \addplot[xbar,fill=black!70, area legend] coordinates {
        (7344*512/953.307/25,1.75)
        (7344*512/740.525/25,3.75)
        };
        \legend {RISC-V, A64FX};
\end{axis}

\end{tikzpicture}
}
\caption{Distributed scaling for a single node and two nodes using MPI for communication on RISC-V and Supercomputer\ Fugaku. Unfortunately, our in-house cluster only had two nodes. Note that we used all cores of the nodes. Recall that each A64FX node has 48 cores and each RISC-V node has 64 cores.}
    \label{fig:performance:distributed}
\end{figure}

Figure~\ref{fig:distributed:11:initial} shows the results for the DWD Separated of Level 11. Here, on the single node both runs are comparable. However, on two nodes the RISC-V run is slightly faster. However, we must consider that an A64FX node has 48 cores and a RISC-V node has 64. Table~\ref{tab:distributed:dwd:flops} shows the MFLOP/s for all levels.

\begin{figure}[tb]
    \centering
  
    \begin{tikzpicture}[scale=0.95, transform shape]
 \begin{axis}[
    xbar=12pt,
    xmin=0,xmax=800,
    ytick=data,
    enlarge y limits={abs=1cm},
    yticklabels={1,1,2,2},
    ytick={1,2,3,4},
    bar width = 10pt,
    xlabel=Processed sub-grids per time step per second, 
    ytick align=outside, 
    ytick pos=left,
    ylabel= \# nodes,
    major x tick style ={ transparent},
    legend style={at={(0.64,0.96)},anchor=north west, font=\footnotesize, legend cell align=left},
    xmajorgrids=true,
    nodes near coords,
    title=Level 11 (Separated)
    ]    
    \addplot[xbar,fill=cadetgrey!20, area legend] coordinates {
        (4796*512/927.38/25,1.25)
        (4796*512/557.73/25,3.25)
        };
        \addplot[xbar,fill=black!70, area legend] coordinates {
        (4796*512/600.244/25,1.75)
        (4796*512/436.271/25,3.75)
        };
        \legend {RISC-V, A64FX};
\end{axis}
\end{tikzpicture}
    \caption{Distributed scaling for a single node and two nodes using MPI for communication on RISC-V and Supercomputer\ Fugaku. Unfortunately, our in-house cluster only had two nodes. Note that we used all cores of the nodes. Recall that an A64FX node has 48 cores and a RISC-V node has 64 cores.}
    \label{fig:distributed:11:initial}
\end{figure}

\begin{table}[tb]
    \centering
    \caption{MFLOP per time step for single node and two node runs on RISC-V and A64FX. Note that a RISC-V node has 64 cores and a A64FX code has 48 cores.}
    \label{tab:distributed:dwd:flops}
    \begin{tabular}{l|ll|ll}\toprule
            & \multicolumn{2}{c}{1 node} &  \multicolumn{2}{c}{2 nodes} \\\midrule
     Level    & RISC-V & A64FX  & RISC-V & A64FX \\ \midrule
     10 (initial) & \num{153.34} & \num{93.56} & \num{196.68} & \num{105.86}\\
     10 (refined) & \num{148.87} & \num{88.11}  & \num{248.57} & \num{113.43}\\
     11 (initial) & \num{161.58} & \num{73.97} & \num{261.67} & \num{101.77}\\\bottomrule
    \end{tabular}
\end{table}

\subsubsection{V1309}
In this section, we look into the scaling of the production run of the v1309 scenario. Here, we ran the code for 10 time steps. Table~\ref{tab:scaling:v1309} shows the distributed scaling. Here, due to the restricted limit of 28 GB of memory per node, we had to use 16 Supercomputer\ Fugaku nodes. We experienced similar issues with less memory on Supercomputer\ Fugaku while comparing with NERSC'S Perlmutter~\cite{diehl-fugaku2024}. We observe that we get around 100 more sub-grids per second processed on RISC-V using a single node as for 16 supercomputer\ Fugaku nodes.

\begin{table}[tb]
    \centering
    \caption{Comparision for v1309 on a single MILK-V with 128 GB node and 16 A64FX nodes with 28 GB per node.}
    \begin{tabular}{l|l}
     \toprule
     Computer & Processed sub-grids per second\\\midrule
     SuperComputer\ Fugaku    & \num{635.7} \\
     MILK-V    &  \num{740.3}\\\bottomrule 
    \end{tabular}
    \label{tab:scaling:v1309}
\end{table}

\section{Power measurements}
\label{sec:power}
We compared the RISC-V boards and Supercomputer\ Fugaku for both scenarios. For the Sophon RISC-V CPU, no hardware counters are available. For the single-board computers, we attached a USB power meter to approximate the power consumption~\cite{10.1145/3457388.3458657}. However, the desktop computer has a regular connector for regular sockets. The manufacturer reports a typical power consumption of 120 watts\footnote{\url{https://milkv.io/pioneer}}. The typical power consumption of one supercomputer\ Fugaku node is around 200 Watts~\cite{sreepathi2021early}. 

\subsubsection{DWD}
\begin{figure}[tb]
    \centering
    \subfloat[\label{fig:performance:power:initial}]{
\begin{tikzpicture}[scale=0.95, transform shape]
 \begin{axis}[
    xbar=12pt,
    xmin=0,xmax=2000,
    ytick=data,
    enlarge y limits={abs=1cm},
    yticklabels={1,1,2,2},
    ytick={1,2,3,4},
    bar width = 10pt,
    xlabel=Wh, 
    ytick align=outside, 
    ytick pos=left,
    ylabel= \# nodes,
    major x tick style ={transparent},
    legend style={at={(0.34,0.96)},anchor=north west, font=\footnotesize, legend cell align=left},
    xmajorgrids=true,
    nodes near coords,
    title=Level 10 (Separated)
    ]    
    \addplot[xbar,fill=cadetgrey!20, area legend] coordinates {
        (120*172.159/60,1.25)
        (2*120*134.226/60,3.25)
        };
        \addplot[xbar,fill=black!70, area legend] coordinates {
        (200*237.02/60,1.75)
        (2*200*209.48/60,3.75)
        };
        \legend {RISC-V, A64FX};
\end{axis}

\end{tikzpicture}
}

\subfloat[\label{fig:performance:power:refined}]{
\begin{tikzpicture}[scale=0.95, transform shape]
 \begin{axis}[
    xbar=12pt,
    xmin=0,xmax=6000,
    ytick=data,
    enlarge y limits={abs=1cm},
    yticklabels={1,1,2,2},
    ytick={1,2,3,4},
    bar width = 10pt,
    xlabel=Wh, 
    ytick align=outside, 
    ytick pos=left,
    ylabel= \# nodes,
    major x tick style ={ transparent},
    legend style={at={(0.34,0.96)},anchor=north west, font=\footnotesize, legend cell align=left},
    xmajorgrids=true,
    nodes near coords,
    title=Level 10 (DWD Merging)
    ]    
    \addplot[xbar,fill=cadetgrey!20, area legend] coordinates {
        (120*564.26/60,1.25)
        (2*120*337.911/60,3.25)
        };
        \addplot[xbar,fill=black!70, area legend] coordinates {
        (200*953.307/60,1.75)
        (2*200*740.525/60,3.75)
        };
        \legend {RISC-V, A64FX};
\end{axis}
\end{tikzpicture}
}
\caption{Power consumption for a single node and two nodes using MPI for communication on RISC-V and Supercomputer\ Fugaku. Unfortunately, our in-house cluster only had two nodes. Note that we used all cores of the nodes. Recall that an A64FX node has 48 cores and a RISC-V node has 64 cores.}
    \label{fig:performance:distributed:power}
\end{figure}
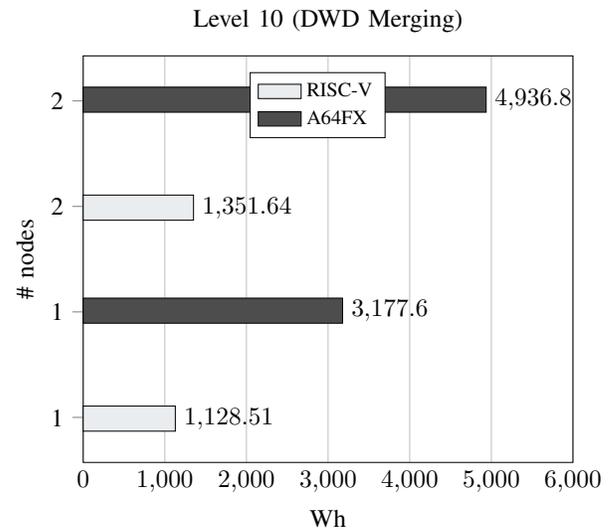

Figure~\ref{fig:performance:distributed:power} shows the power consumption for both meshes of level 10. Here, RISC-V has around two times less power consumption on a single node and three times less on two nodes.

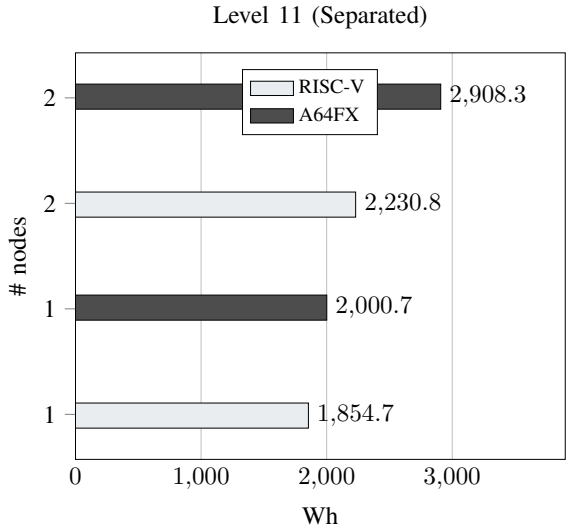
\begin{figure}[tb]
    \centering
    \begin{tikzpicture}[scale=0.95, transform shape]
    \begin{axis}[
    xbar=12pt,
    xmin=0,xmax=3900,
    ytick=data,
    enlarge y limits={abs=1cm},
    yticklabels={1,1,2,2},
    ytick={1,2,3,4},
    bar width = 10pt,
    xlabel=Wh, 
    ytick align=outside, 
    ytick pos=left,
    ylabel= \# nodes,
    major x tick style ={ transparent},
    legend style={at={(0.34,0.96)},anchor=north west, font=\footnotesize, legend cell align=left},
    xmajorgrids=true,
    nodes near coords,
    title=Level 11 (Separated)
    ]    
    \addplot[xbar,fill=cadetgrey!20, area legend] coordinates {
        (120*927.38/60,1.25)
        (2*120*557.73/60,3.25)
        };
        \addplot[xbar,fill=black!70, area legend] coordinates {
        (200*600.244/60,1.75)
        (2*200*436.271/60,3.75)
        };
        \legend {RISC-V, A64FX};
\end{axis}
\end{tikzpicture}
    \caption{Power consumption for a single node and two nodes using MPI for communication on RISC-V and Supercomputer\ Fugaku. Unfortunately, our in-house cluster only had two nodes. Note that we used all cores of the nodes. Recall that an A64FX node has 48 cores and a RISC-V node has 64 cores.}
    \label{fig:distributed:11:power}
\end{figure}

Figure~\ref{fig:distributed:11:power} shows the power consumption for the DWD Separated of level 11. Here, both architectures have a similar power consumption. Note that on supercomputer\ Fugaku Sandia's PowerAPI~\cite{grant2016standardizing} is integrated with the PJM job scheduler. Thus, for all runs, information about the power consumption is available. For the used RISC-V CPU, no hardware counters or software tools are available yet. For future comparison, it would be beneficial to obtain a more accurate power consumption measurement on RISC-V. For a fair comparison in this paper, we used the typical energy consumption for a single node provided by the manufacturer.

\subsubsection{V1309}
Table~\ref{tab:power:v1309} shows the power consumption for a single time step of the v1309 scenario. The scenario fitted in one RISC-V node. Due to the 28 GB memory of the supercomputer\ Fugaku nodes, we had to use 16 nodes. The power consumption is around 50\% higher on A64FX.

\begin{table}[tb]
    \centering
    \caption{Comparision for v1309 on a single MILK-V with 128 GB node and 16 A64FX nodes with 28 GB per node.}
    \begin{tabular}{l|l}
     \toprule
     Computer & Wh\\\midrule
     SuperComputer\ Fugaku    & \num{57831} \\
     MILK-V    &  \num{29798}\\\bottomrule 
    \end{tabular}
    \label{tab:power:v1309}
\end{table}

\section{Conclusion}
\label{sec:conclusion}
The Pioneer RISC-V machines used in this paper are the first desktop-grade systems available for general use.
While most of the porting of the HPX to RISC-V was done in the author's previous work~\cite{diehl2023evaluating}, this study required additional optimizations for HPX's atomic support for Arm and RISC-V. Furthermore, to explore the vectorization the authors developed the RISC-V vector library. In addition, changes were made to Kokkos' CMake build system in order to support the Pioneer's RISC-V Sophon CPU. Modifications to Kokkos' CMake system brought to the forefront questions about how to handle different flavors of RISC-V processors in the Kokkos build system moving forward.

Previously, we compared Octo-Tiger's performance on single-board computers with 4 cores and memory controllers which did not meet the expectations for HPC-grade hardware. Now, with 64 cores and faster memory controllers, larger scenarios
can be investigated. In addition, we investigated the effect of vectorization on an 8-core single-board computer since this was the only available option with a RISC-V CPU with full vectorization support. We observed improvements using vectorization. However, we would like to investigate vectorization on desktop-grade hardware.
For the rotating star scenario, A64FX and RISC-V showed comparable performance with the improved HPX support for atomics on A64FX. The improved atomics showed improvements on RISC-V for larger scenarios too. We observed scaling from a single core up to 64 cores on RISC-V for the node-level runs. For a detailed node-level performance analysis on A64FX for Ookami and supercomputer\ Fugaku, we refer to~\cite{diehl-fugaku2024}.

For the first scenario, the double white dwarf (DWD), we compared the sub-grids processed per time step per second and the floating point operations per time step on a single node and two nodes. In most cases, the RISC-V nodes were slightly faster. One reason for the performance difference can be attributed to the A64FX nodes have 48 cores and the RISC-V nodes have 64 cores. For the second scenario, the v1309, the input file was much larger and the scenario did not fit into a single A64FX node due to the 28 GB memory limit per node.
Instead, 16 nodes were required.
Here, we got comparable sub-grids processed per time step per second. Using the desktop-grade RISC-V hardware resulted in similar performance concerning A64FX taking the core difference into account.

When considering power consumption, we observed the RISC-V system generally used less energy for some runs. The typical energy consumption of a RISC-V node is 120 Watts, and for a supercomputer\ Fugaku node is 200 Watts~\cite{sreepathi2021early}. In a couple of instances, however, both architectures shared comparable power utilization.

Octo-Tiger supports GPU's via Kokkos. We believe a follow-up study of heterogeneous computations using RISC-V CPUs and GPUs is supported by this work. As of this writing, there are no CUDA drivers or NVIDIA SDK for RISC-V. AMD's latest consumer GPUs work on RISC-V platforms. In light of AMD's support for RISC-V, a follow-up study investigating Kokkos kernel execution on a heterogeneous system with an AMD GPU and RISC-V CPU is worth pursuing. All the steps in our study are in preparation for the European Processor Initiative (EPI). The European Union has recently announced a substantial investment towards the development of HPC RISC-V hardware and software.  These efforts are pivotal in laying the groundwork for the eventual deployment of the RISC-V-based European supercomputer slated for launch in 2026.    

Overall, this work demonstrates the viability of current RISC-V hardware generation for real-world astrophysics simulations. Notably, we encountered no significant issues running Octo-Tiger on this first-gen RISC-V desktop hardware. Even given Octo-Tiger's somewhat unusual software stack (using HPX over MPI+OpenMP), production-grade simulations easily scaled to all cores and multiple nodes on this novel hardware platform.

Overall, our experiences give us cautious optimism regarding the platform itself. We plan to repeat this benchmark once newer RISC-V (server-grade) platforms are available. Looking toward future RISC-V studies, several vendors and opportunities are on the horizon. We look forward to assessing Octo-Tiger's performance on the following RISC-V architectures: Ventana's Veyron V2, X-Silicon's C-GPU, and Tenstorrent's Ascalon. Since the European RISC-V supercomputer will be a CPU-only machine, we have focused on the comparison with supercomputer\ Fugaku which is also a CPU-only machine.

The following features added to the RISC-V ISA would be beneficial for asynchronous many-task runtime systems like HPX: one-cycle context switches, extended atomics, hardware support for global address space, and hardware support for thread scheduling (hardware queues). For general HPC workloads the RISC-V vector machine extension, user-land cache management and control, and hardware counters for power consumption would be beneficial.

\section*{Acknowledgment}
{\footnotesize
This research employed computational resources of the supercomputer Fugaku provided by RIKEN Center for Computational Science. We thank the LSU Center of Computation \& Technology for supporting this work. The authors would like to thank Stony Brook Research Computing and Cyberinfrastructure, and the Institute for Advanced Computational Science at Stony Brook University for access to the innovative high-performance Ookami computing system, which was made possible by a \$5M National Science Foundation grant (\#1927880). This work was supported by the U.S. Department of Energy through the Los Alamos National Laboratory. Los Alamos National Laboratory is operated by Triad National Security, LLC, for the National Nuclear Security Administration of U.S. Department of Energy (Contract No. 89233218CNA000001). LA-UR-24-28599}

\section*{Supplementary materials}
{\footnotesize
The build scripts to build Octo-Tiger are available on GitHub\footnote{\url{https://github.com/STEllAR-GROUP/OctoTigerBuildChain}}. Octo-Tiger\footnote{\url{https://github.com/STEllAR-GROUP/octotiger}}, HPX\footnote{\url{https://github.com/STEllAR-GROUP/hpx}}, and HPX-Kokkos\footnote{\url{https://github.com/STEllAR-GROUP/hpx-kokkos}} are available on GitHub. The input files for the v1309 scenario are available on Zenodo\footnote{https://doi.org/10.5281/zenodo.5213015}. The slurm job scripts and specific input data are available on GitHub\footnote{\url{https://github.com/diehlpkpapers/RISC-V-24}}.
}

\bibliographystyle{IEEEtran}
\bibliography{references}

\begin{thebibliography}{10}
\providecommand{\url}[1]{#1}
\csname url@samestyle\endcsname
\providecommand{\newblock}{\relax}
\providecommand{\bibinfo}[2]{#2}
\providecommand{\BIBentrySTDinterwordspacing}{\spaceskip=0pt\relax}
\providecommand{\BIBentryALTinterwordstretchfactor}{4}
\providecommand{\BIBentryALTinterwordspacing}{\spaceskip=\fontdimen2\font plus
\BIBentryALTinterwordstretchfactor\fontdimen3\font minus \fontdimen4\font\relax}
\providecommand{\BIBforeignlanguage}[2]{{%
\expandafter\ifx\csname l@#1\endcsname\relax
\typeout{** WARNING: IEEEtran.bst: No hyphenation pattern has been}%
\typeout{** loaded for the language `#1'. Using the pattern for}%
\typeout{** the default language instead.}%
\else
\language=\csname l@#1\endcsname
\fi
#2}}
\providecommand{\BIBdecl}{\relax}
\BIBdecl

\bibitem{waterman2014risc}
A.~Waterman, Y.~Lee, D.~Patterson, K.~Asanovic, V.~I.~U. level Isa, A.~Waterman, Y.~Lee, and D.~Patterson, ``{The RISC-V instruction set manual},'' \emph{Volume I: User-Level ISA’, version}, vol.~2, pp. 1--79, 2014.

\bibitem{blind2021impact}
K.~Blind, M.~B{\"o}hm, P.~Grzegorzewska, A.~Katz, S.~Muto, S.~P{\"a}tsch, and T.~Schubert, ``{The impact of Open Source Software and Hardware on technological independence, competitiveness and innovation in the EU economy},'' \emph{Final Study Report. European Commission, Brussels, doi}, vol.~10, p. 430161, 2021.

\bibitem{kovavc2019european}
M.~Kova{\v{c}} \emph{et~al.}, ``{European processor initiative (EPI)—An approach for a future automotive eHPC semiconductor platform},'' in \emph{Electronic Components and Systems for Automotive Applications: Proceedings of the 5th CESA Automotive Electronics Congress, Paris, 2018}.\hskip 1em plus 0.5em minus 0.4em\relax Springer, 2019, pp. 185--195.

\bibitem{10.1145/3457388.3458657}
\BIBentryALTinterwordspacing
A.~D\"{o}rflinger \emph{et~al.}, ``{A Comparative Survey of Open-Source Application-Class RISC-V Processor Implementations},'' in \emph{Proceedings of the 18th ACM International Conference on Computing Frontiers}, ser. CF '21.\hskip 1em plus 0.5em minus 0.4em\relax New York, NY, USA: Association for Computing Machinery, 2021, p. 12–20. [Online]. Available: \url{https://doi.org/10.1145/3457388.3458657}
\BIBentrySTDinterwordspacing

\bibitem{diehl2023evaluating}
P.~Diehl, G.~Daiss, S.~Brandt, A.~Kheirkhahan, H.~Kaiser, C.~Taylor, and J.~Leidel, ``{Evaluating HPX and Kokkos on RISC-V using an astrophysics application Octo-Tiger},'' in \emph{Proceedings of the SC'23 Workshops of The International Conference on High Performance Computing, Network, Storage, and Analysis}, 2023, pp. 1533--1542.

\bibitem{hornung2017raja}
R.~D. Hornung and H.~E. Hones, ``Raja performance suite,'' Lawrence Livermore National Lab.(LLNL), Livermore, CA (United States), Tech. Rep., 2017.

\bibitem{10.1145/3624062.3624234}
\BIBentryALTinterwordspacing
N.~Brown, M.~Jamieson, J.~Lee, and P.~Wang, ``{Is RISC-V ready for HPC prime-time: Evaluating the 64-core Sophon SG2042 RISC-V CPU},'' in \emph{Proceedings of the SC '23 Workshops of The International Conference on High Performance Computing, Network, Storage, and Analysis}, ser. SC-W '23.\hskip 1em plus 0.5em minus 0.4em\relax New York, NY, USA: Association for Computing Machinery, 2023, p. 1566–1574. [Online]. Available: \url{https://doi.org/10.1145/3624062.3624234}
\BIBentrySTDinterwordspacing

\bibitem{9771410}
B.~W. Mezger, D.~A. Santos, L.~Dilillo, C.~A. Zeferino, and D.~R. Melo, ``{A Survey of the RISC-V Architecture Software Support},'' \emph{IEEE Access}, vol.~10, pp. 51\,394--51\,411, 2022.

\bibitem{stoffelen2019efficient}
K.~Stoffelen, ``{Efficient cryptography on the RISC-V architecture},'' in \emph{Progress in Cryptology--LATINCRYPT 2019: 6th International Conference on Cryptology and Information Security in Latin America, Santiago de Chile, Chile, October 2--4, 2019, Proceedings 6}, Springer.\hskip 1em plus 0.5em minus 0.4em\relax Santiago de Chile, Chile: Springer, 2019, pp. 323--340.

\bibitem{louis2019towards}
M.~S. Louis \emph{et~al.}, ``{Towards deep learning using tensorflow lite on RISC-V},'' in \emph{Third Workshop on Computer Architecture Research with RISC-V (CARRV)}, vol.~1.\hskip 1em plus 0.5em minus 0.4em\relax Phoenix, Arizona: ACM, 2019, p.~6.

\bibitem{schiavone2017slow}
P.~D. Schiavone \emph{et~al.}, ``{Slow and steady wins the race? A comparison of ultra-low-power RISC-V cores for Internet-of-Things applications},'' in \emph{2017 27th International Symposium on Power and Timing Modeling, Optimization and Simulation (PATMOS)}, IEEE.\hskip 1em plus 0.5em minus 0.4em\relax Thessaloniki, Greece: IEEE, 2017, pp. 1--8.

\bibitem{Kaiser2020}
\BIBentryALTinterwordspacing
H.~Kaiser, P.~Diehl, A.~S. Lemoine, B.~A. Lelbach, P.~Amini, A.~Berge, J.~Biddiscombe, S.~R. Brandt, N.~Gupta, T.~Heller, K.~Huck, Z.~Khatami, A.~Kheirkhahan, A.~Reverdell, S.~Shirzad, M.~Simberg, B.~Wagle, W.~Wei, and T.~Zhang, ``{HPX - The C++ Standard Library for Parallelism and Concurrency},'' \emph{Journal of Open Source Software}, vol.~5, no.~53, p. 2352, 2020. [Online]. Available: \url{https://doi.org/10.21105/joss.02352}
\BIBentrySTDinterwordspacing

\bibitem{10.1145/3624062.3624598}
\BIBentryALTinterwordspacing
J.~Yan, H.~Kaiser, and M.~Snir, ``{Design and Analysis of the Network Software Stack of an Asynchronous Many-task System -- The LCI parcelport of HPX},'' in \emph{Proceedings of the SC '23 Workshops of The International Conference on High Performance Computing, Network, Storage, and Analysis}, ser. SC-W '23.\hskip 1em plus 0.5em minus 0.4em\relax New York, NY, USA: Association for Computing Machinery, 2023, p. 1151–1161. [Online]. Available: \url{https://doi.org/10.1145/3624062.3624598}
\BIBentrySTDinterwordspacing

\bibitem{10.1145/3295500.3356221}
\BIBentryALTinterwordspacing
G.~Dai\ss{} \emph{et~al.}, ``{From Piz Daint to the Stars: Simulation of Stellar Mergers Using High-Level Abstractions},'' in \emph{{Proceedings of the International Conference for High Performance Computing, Networking, Storage and Analysis}}, ser. SC '19.\hskip 1em plus 0.5em minus 0.4em\relax New York, NY, USA: Association for Computing Machinery, 2019. [Online]. Available: \url{https://doi.org/10.1145/3295500.3356221}
\BIBentrySTDinterwordspacing

\bibitem{9556040}
P.~Diehl, G.~Daiß, D.~Marcello, K.~Huck, S.~Shiber, H.~Kaiser, J.~Frank, G.~C. Clayton, and D.~Pflüger, ``{Octo-Tiger’s New Hydro Module and Performance Using HPX+CUDA on ORNL’s Summit},'' in \emph{2021 IEEE International Conference on Cluster Computing (CLUSTER)}, 2021, pp. 204--214.

\bibitem{diehl-fugaku2024}
\BIBentryALTinterwordspacing
P.~Diehl, G.~Dai{\ss}, K.~Huck, D.~Marcello, S.~Shiber, H.~Kaiser, and D.~Pfl{\"u}ger, ``{Simulating stellar merger using HPX/Kokkos on A64FX on Supercomputer Fugaku},'' \emph{The Journal of Supercomputing}, 2024. [Online]. Available: \url{https://doi.org/10.1007/s11227-024-06113-w}
\BIBentrySTDinterwordspacing

\bibitem{9485033}
C.~R. Trott, D.~Lebrun-Grandié, D.~Arndt, J.~Ciesko, V.~Dang, N.~Ellingwood, R.~Gayatri, E.~Harvey, D.~S. Hollman, D.~Ibanez, N.~Liber, J.~Madsen, J.~Miles, D.~Poliakoff, A.~Powell, S.~Rajamanickam, M.~Simberg, D.~Sunderland, B.~Turcksin, and J.~Wilke, ``{Kokkos 3: Programming Model Extensions for the Exascale Era},'' \emph{IEEE Transactions on Parallel and Distributed Systems}, vol.~33, no.~4, pp. 805--817, 2022.

\bibitem{9460406}
G.~Daiß, M.~Simberg, A.~Reverdell, J.~Biddiscombe, T.~Pollinger, H.~Kaiser, and D.~Pflüger, ``{Beyond Fork-Join: Integration of Performance Portable Kokkos Kernels with HPX},'' in \emph{2021 IEEE International Parallel and Distributed Processing Symposium Workshops (IPDPSW)}, 2021, pp. 377--386.

\bibitem{10.1145/3585341.3585354}
\BIBentryALTinterwordspacing
G.~Dai\ss{}, P.~Diehl, H.~Kaiser, and D.~Pfl\"{u}ger, ``{Stellar Mergers with HPX-Kokkos and SYCL: Methods of Using an Asynchronous Many-Task Runtime System with SYCL},'' in \emph{Proceedings of the 2023 International Workshop on OpenCL}, ser. IWOCL '23.\hskip 1em plus 0.5em minus 0.4em\relax New York, NY, USA: Association for Computing Machinery, 2023. [Online]. Available: \url{https://doi.org/10.1145/3585341.3585354}
\BIBentrySTDinterwordspacing

\bibitem{daiss2022aggregation}
\BIBentryALTinterwordspacing
G.~Daiß, P.~Diehl, D.~Marcello, A.~Kheirkhahan, H.~Kaiser, and D.~Pflüger, ``{From Task-Based GPU Work Aggregation to Stellar Mergers: Turning Fine-Grained CPU Tasks into Portable GPU Kernels},'' in \emph{2022 IEEE/ACM International Workshop on Performance, Portability and Productivity in HPC (P3HPC)}.\hskip 1em plus 0.5em minus 0.4em\relax Los Alamitos, CA, USA: IEEE Computer Society, nov 2022, pp. 89--99. [Online]. Available: \url{https://doi.ieeecomputersociety.org/10.1109/P3HPC56579.2022.00014}
\BIBentrySTDinterwordspacing

\bibitem{daiss2022simd}
\BIBentryALTinterwordspacing
G.~Daiß, S.~Singanaboina, P.~Diehl, H.~Kaiser, and D.~Pflüger, ``{From Merging Frameworks to Merging Stars: Experiences using HPX, Kokkos and SIMD Types},'' in \emph{2022 IEEE/ACM 7th International Workshop on Extreme Scale Programming Models and Middleware (ESPM2)}.\hskip 1em plus 0.5em minus 0.4em\relax Los Alamitos, CA, USA: IEEE Computer Society, nov 2022, pp. 10--19. [Online]. Available: \url{https://doi.ieeecomputersociety.org/10.1109/ESPM256814.2022.00007}
\BIBentrySTDinterwordspacing

\bibitem{10.1093/mnras/stab937}
\BIBentryALTinterwordspacing
D.~C. Marcello, S.~Shiber, O.~De~Marco, J.~Frank, G.~C. Clayton, P.~M. Motl, P.~Diehl, and H.~Kaiser, ``{octo-tiger: a new, 3D hydrodynamic code for stellar mergers that uses HPX parallelization},'' \emph{Monthly Notices of the Royal Astronomical Society}, vol. 504, no.~4, pp. 5345--5382, 04 2021. [Online]. Available: \url{https://doi.org/10.1093/mnras/stab937}
\BIBentrySTDinterwordspacing

\bibitem{shiber2024hydrodynamic}
S.~Shiber, O.~D. Marco, P.~M. Motl, B.~Munson, D.~C. Marcello, J.~Frank, P.~Diehl, G.~C. Clayton, B.~N. Skinner, H.~Kaiser, G.~Daiss, D.~Pfluger, and J.~E. Staff, ``{Hydrodynamic simulations of WD-WD mergers and the origin of RCB stars},'' 2024.

\bibitem{diehletal2021}
P.~Diehl, D.~Marcello, P.~Amini, H.~Kaiser, S.~Shiber, G.~C. Clayton, J.~Frank, G.~Dais, D.~Pfluger, D.~Eder, A.~Koniges, and K.~Huck, ``{Performance Measurements Within Asynchronous Task-Based Runtime Systems: A Double White Dwarf Merger as an Application},'' \emph{Computing in Science \& Engineering}, vol.~23, no.~03, pp. 73--81, may 2021.

\bibitem{Tylenda_2011}
\BIBentryALTinterwordspacing
R.~Tylenda, M.~Hajduk, T.~Kamiński, A.~Udalski, I.~Soszyński, M.~K. Szymański, M.~Kubiak, G.~Pietrzyński, R.~Poleski, L.~Wyrzykowski, and K.~Ulaczyk, ``{V1309 Scorpii: merger of a contact binary},'' \emph{Astronomy \& Astrophysics}, vol. 528, p. A114, Mar. 2011. [Online]. Available: \url{http://dx.doi.org/10.1051/0004-6361/201016221}
\BIBentrySTDinterwordspacing

\bibitem{Even_2009}
\BIBentryALTinterwordspacing
W.~Even and J.~E. Tohline, ``{CONSTRUCTING SYNCHRONOUSLY ROTATING DOUBLE WHITE DWARF BINARIES},'' \emph{The Astrophysical Journal Supplement Series}, vol. 184, no.~2, p. 248, sep 2009. [Online]. Available: \url{https://dx.doi.org/10.1088/0067-0049/184/2/248}
\BIBentrySTDinterwordspacing

\bibitem{brooks1981dynamics}
R.~Brooks and J.~P. Matelski, ``The dynamics of 2-generator subgroups of psl (2, c),'' in \emph{Riemann surfaces and related topics: Proceedings of the 1978 Stony Brook Conference}, vol.~1.\hskip 1em plus 0.5em minus 0.4em\relax Princeton University Press Princeton, New Jersey, 1981.

\bibitem{Diehl2024}
\BIBentryALTinterwordspacing
P.~Diehl, S.~R. Brandt, and H.~Kaiser, ``{Shared Memory Parallelism in Modern C++ and HPX},'' \emph{SN Computer Science}, vol.~5, no.~5, p. 459, Apr 2024. [Online]. Available: \url{https://doi.org/10.1007/s42979-024-02769-6}
\BIBentrySTDinterwordspacing

\bibitem{patrick_diehl_2023_8067170}
P.~Diehl and C.~Taylor, ``Shared memory parallelism in modern c++,'' https://doi.org/10.5281/zenodo.8067170, Jun. 2023.

\bibitem{diehl2023simulating}
P.~Diehl, G.~Dai{\ss}, K.~Huck, D.~Marcello, S.~Shiber, H.~Kaiser, and D.~Pfl{\"u}ger, ``{Simulating Stellar Merger using HPX/Kokkos on A64FX on Supercomputer Fugaku},'' in \emph{2023 IEEE International Parallel and Distributed Processing Symposium Workshops (IPDPSW)}.\hskip 1em plus 0.5em minus 0.4em\relax IEEE, 2023, pp. 682--691.

\bibitem{sreepathi2021early}
S.~Sreepathi and M.~Taylor, ``{Early Evaluation of Fugaku A64FX Architecture Using Climate Workloads},'' in \emph{2021 IEEE International Conference on Cluster Computing (CLUSTER)}.\hskip 1em plus 0.5em minus 0.4em\relax IEEE, 2021, pp. 719--727.

\bibitem{grant2016standardizing}
R.~E. Grant, M.~Levenhagen, S.~L. Olivier, D.~DeBonis, K.~T. Pedretti, and J.~H.~L. III, ``{Standardizing Power Monitoring and Control at Exascale},'' \emph{Computer}, vol.~49, no.~10, pp. 38--46, Oct 2016.

\bibitem{marcello_2021_5213015}
\BIBentryALTinterwordspacing
Marcello, Shiber, and Diehl, ``Restart data for the v1309 scenario,'' Aug. 2021. [Online]. Available: \url{https://doi.org/10.5281/zenodo.5213015}
\BIBentrySTDinterwordspacing

\end{thebibliography}

\end{document}